\documentclass[conference]{IEEEtran}
\IEEEoverridecommandlockouts

\newcommand{\linebreakand}{%
  \end{@IEEEauthorhalign}
  \hfill\mbox{}\par
  \mbox{}\hfill\begin{@IEEEauthorhalign}
}

\usepackage[english]{babel}

\usepackage[letterpaper,top=2cm,bottom=2cm,left=3cm,right=3cm,marginparwidth=1.75cm]{geometry}

\usepackage{amsmath}
\usepackage{graphicx}
\usepackage[colorlinks=true, allcolors=blue]{hyperref}
\usepackage{hyperref}

\begin{document}

\title{Cellular Segmentation and Composition in Routine Histology Images using Deep Learning}

\author{
\IEEEauthorblockN{Muhammad Dawood}
\IEEEauthorblockA{\textit{Department of Computer Science} \\
\textit{University of Warwick}\\
Coventry, United Kingdom \\
muhammad.dawood@warwick.ac.uk}
\and

\IEEEauthorblockN{Raja Muhammad Saad Bashir}
\IEEEauthorblockA{\textit{Department of Computer Science} \\
\textit{University of Warwick}\\
Coventry, United Kingdom \\
saad.bashir@warwick.ac.uk}
\and

\IEEEauthorblockN{Srijay Deshpande}
\IEEEauthorblockA{\textit{Department of Computer Science} \\
\textit{University of Warwick}\\
Coventry, United Kingdom \\
srijay.deshpande@warwick.ac.uk}
\and

\IEEEauthorblockN{Manahil Raza}
\IEEEauthorblockA{\textit{Department of Computer Science} \\
\textit{University of Warwick}\\
Coventry, United Kingdom \\
manahil.raza@warwick.ac.uk}
\and 
\IEEEauthorblockN{Adam Shephard}
\IEEEauthorblockA{\textit{Department of Computer Science} \\
\textit{University of Warwick}\\
Coventry, United Kingdom \\
adam.shephard@warwick.ac.uk}
}

\maketitle

\begin{abstract}
Identification and quantification of nuclei in  colorectal cancer haematoxylin \& eosin (H\&E) stained histology images is crucial to prognosis and patient management. In computational pathology these tasks are referred to as nuclear segmentation, classification and composition and are used to extract meaningful interpretable cytological and architectural features for downstream analysis. The CoNIC challenge poses the task of automated nuclei segmentation, classification and composition into six different types of nuclei from the largest publicly known nuclei dataset - Lizard. In this regard, we have developed pipelines for the prediction of nuclei segmentation using HoVer-Net and ALBRT for cellular composition. On testing on the preliminary test set, HoVer-Net achieved a PQ of 0.58, a PQ+ of 0.58 and finally a mPQ+ of 0.35. For the prediction of cellular composition with ALBRT on the preliminary test set, we achieved an overall $R^2$ score of 0.53, consisting of 0.84 for lymphocytes, 0.70 for epithelial cells, 0.70 for plasma and .060 for eosinophils.
\end{abstract}

\begin{IEEEkeywords}
Segmentation, Pathology, Deep Learning, Cellular Composition
\end{IEEEkeywords}

\section{Introduction}
Colorectal cancer starts from the colon/rectum that makes up the large intestine in the digestive system. There are many types of colorectal cancers, with the most common being adenocarcinomas, and others such as carcinoid tumours being more rare. TNM (Tumor, Node and Metastasis) staging is used for the detection and diagnosis of colorectal cancer ranging from 0 (earliest stage) to IV (latest stage) \cite{acs}. The segmentation of malignant nuclei in the colon wall may provide important predictive features for the correct staging of colorectal cancers. Thus, the CoNIC  \cite{graham2021conic} challenge was curated. This challenge encourages contestants to develop algorithms to perform segmentation, classification and counting of six different nuclei types using the largest publicly available nuclei dataset - Lizard \cite{graham2021lizard}. This import processing step may enable future researchers to improve automated cancer staging, whilst using interpretable features. In this regard, as part of the CoNIC Challenge, we have performed two tasks as a part of the \textbf{TIA Warwick} team: 1. Nuclear segmentation and classification using HoVer-Net \cite{graham2019hover} and 2. Prediction of cellular composition, using a modified version of ALBRT \cite{dawood2021albrt}. We describe our pipeline below.

\section{Nuclear Segmentation and Classification}

We utilized the HoVer-Net \cite{graham2019hover} framework to perform simultaneous segmentation and classification of nuclear instances on the histology images from the CoNIC dataset. The classification task characterizes the detected nuclei into one of the following cell types:  epithelial, lymphocyte, plasma, eosinophil, neutrophil or connective tissue cells. 

HoVer-Net is a deep learning framework that
consists of an encoder branch, and three decoder branches. The encoder branch is based on the pre-activated ResNet-50 \cite{he2016identity}. The decoder branches then perform up-sampling to achieve segmentation and classification. There are three decoder branches: the nuclear pixel branch, the HoVer branch and the nuclear classification branch. The first two branches aim to detect and segment nuclei, with the HoVer branch improving segmentation quality in the presence of overlapping/touching nuclei. Finally, the nuclear classification branch provides a pixel-wise nuclear classification score. Following post-processing (see \cite{graham2019hover}), the maximum argument from the nuclear classification branch is used to obtain the class for each nuclear instance detected by the other two branches.

The available dataset \cite{graham2021lizard} was already partitioned into patches of 256$\times$256  at 20$\times$ objective power by the challenge organisers, which we then utilised in our experiments. We further split the dataset into training and validation sets in a stratified manner (according to the centre from which the image was scanned at). The encoder branch of the HoVer-Net model was pre-trained on the ImageNet weights from the pre-activated ResNet-50 \cite{he2016identity} model. We performed optimisation on the training/validation set, using standard augmentation techniques. The following model hyper-parameters were found through optimisation on the training/validation set: learning rate = 0.001, number of epochs = 50 and batch size of 5. On validation, HoVer-Net achieved a PQ of 0.48 and mPQ+ of 0.40.

Following optimisation, HoVer-Net was then tested on the unseen data provided by the organisers. For this we achieved a PQ of 0.58, a PQ+ of 0.58 and a mPQ+ of 0.35. We also obtained component-level PQ+s of 0.38 (connective tissue), 0.38 (eosinophils), 0.05 (epithelial cells), 0.55 (lymphocytes), 0.35 (neutrophils), 0.40 (plasma). The model performed well on the lymphocytes, plasma, connective tissue etc., but didn't perform as well on the epithelial cells.

\section{Prediction of Cellular Composition}

\begin{table}
\caption{The mean (standard deviation) performance of the modified ALBRT for predicting the counts of different types of cells in term of Coefficient of determination ($R^2$), Mean Absolute Error (MAE), and Mean Arctangent Absolute Percentage Error (MAAPE). \label{regression_results}}
\centering
\begin{tabular}{llll}
\hline
\textbf{Component} & \textbf{R\textsuperscript{2}}  & \textbf{MAE}  & \textbf{MAAPE} \\ \hline
Neutrophil          &   0.17     &   3.16   &  0.32   \\
Epithelial          &   0.70     &   10.15  &  0.25    \\
Lymphocyte          &   0.84     &   2.70   &  0.43       \\
Plasma              &   0.70     &   2.95   &  0.51  \\
Connective tissue   &   0.16     &   6.07   &  0.65  \\
Eosinophil          &   0.59     &   0.44   &  0.14 \\ \hline
  All               &  0.53      &  4.24    &  0.38 \\\hline
\end{tabular}
\end{table}

For the second task, we used ALBRT \cite{dawood2021albrt}, a framework for cellular composition prediction in routine histology images. The original paper takes an input image at 40$\times$ resolution and generates sub-patch level cellular counts. However, in this work the input images are at 20$\times$ resolution. Computing sub-patch level cellular counts, as used by the original ALBRT paper, at this magnification may be noisy, owing to most of the nuclei lying at the center of the patch. Therefore, we only used the backbone Xception network \cite{chollet2017xception} of ALBRT for predicting the counts of different types of cells. This branch takes the full patch as input, and doesn't use sub-patches, thus eliminating this noise issue. We additionally changed the loss function from a ranking loss to Huber loss \cite{huber1992robust} as the method aims to directly maximise the $R^2$ score. We used patch level cellular counts information of six cell types (neuotropil, epithelial, lymphocyte, plasma, eosinophil, connective tissue) using the ground truth instance segmentation mask. During training we shift the image randomly by 10 pixels so that the nuclei at the border of input patch are ignored as the ground truth cellular counts were obtained by taking a window of 224$\times$224 from the patch center.

Since there are very few eosinophils present in the dataset, we trained two sets of models for this work. The first set of models predicted cell counts for all cell types, excluding eosinophils. On 5-fold cross-validation we achieved a R\textsuperscript{2} score of 0.36 (neutrophils), 0.94 (epithelial cells), 0.94 (lymphocytes), 0.74 (plasma) and 0.85 (connective tissue). We then additionally trained a separate modified ALBRT model for predicting eosinophil cell counts alone. On validation this achieved a R\textsuperscript{2} score of 0.76. For the preliminary test set, we used the first set of trained networks to predict the cell count for all cells types, excluding eosinophils. We then averaged their prediction to generate final cell counts for these cell types. The same process was used for the models trained to predict eosinophil counts alone. We achieved an overall $R^2$ of 0.53. For all cell types, other than neutrophils the model has shown good performance on the preliminary test set as shown in Table \ref{regression_results}. This may be a result of there being very few neutrophils present in the dataset.

\section{Conclusion}
In this challenge we  have used HoVer-Net and ALBRT for the segmentation, classification and composition of six different types of nuclei from the H\&E histology images. Both of them have shown competitive performances in the challenge preliminary test sets by achieving position in top 10 positions. We further plan to extend this to extensively test the performance of these algorithms on the full test set released by the CoNIC challenge.

\newpage

\bibliographystyle{plain}
\bibliography{ms.bib}
\end{document}